# Artificial Intelligence Quotient (AIQ): A Novel Framework for Measuring Human-AI Collaborative Intelligence

Venkat Ram Reddy Ganuthula[1]     Krishna Kumar Balaraman[1]

[1] Indian Institute of Technology Jodhpur

## Abstract

As artificial intelligence becomes increasingly integrated into professional and personal domains, traditional metrics of human intelligence require reconceptualization. This paper introduces the Artificial Intelligence Quotient (AIQ), a novel measurement framework designed to assess an individual's capacity to effectively collaborate with and leverage AI systems, particularly Large Language Models (LLMs). Building upon established cognitive assessment methodologies and contemporary AI interaction research, we present a comprehensive framework for quantifying human-AI collaborative intelligence. This work addresses the growing need for standardized evaluation of AI-augmented cognitive capabilities in educational and professional contexts.

## 1. Introduction

The emergence of artificial intelligence as a ubiquitous cognitive tool marks a fundamental shift in human intellectual work. As AI systems become increasingly sophisticated and accessible, the ability to effectively collaborate with these systems emerges as a crucial determinant of professional and educational success. This transformation challenges traditional notions of human intelligence and creates an urgent need for new frameworks to understand and measure human-AI collaborative capabilities.

Traditional intelligence measures, developed in the early 20th century and refined over decades, were designed to assess cognitive capabilities in a pre-AI world. As Sternberg (2019) argues, these measures fail to capture the unique cognitive demands of AI-augmented work environments. The rapid integration of AI into professional, educational, and personal domains creates what Brynjolfsson and McAfee (2017) term a "second machine age," where human success increasingly depends on the ability to effectively leverage AI capabilities.

The democratization of AI access through large language models and other AI tools has created what Howard and Borenstein (2018) describe as a "cognitive augmentation revolution." This revolution transforms not only how we work but how we think and solve problems. However, as Hernández-Orallo (2019) notes, we lack systematic ways to evaluate individual capabilities in this new paradigm. This gap creates significant challenges for educational institutions, organizations, and individuals attempting to navigate the transition to an AI-augmented future.

This paper introduces the concept of Artificial Intelligence Quotient (AIQ) as a framework for understanding and measuring human-AI collaborative capabilities. We argue that AIQ represents a distinct and measurable form of intelligence that will increasingly determine success in professional and educational contexts. Through systematic analysis of current research and emerging workplace demands, we present the case for developing standardized AIQ assessment methodologies.

## 2. Current Approaches and Their Limitations

### 2.1 Traditional Intelligence Assessment

Traditional approaches to measuring human cognitive capabilities, while valuable in their historical context, show significant limitations in addressing AI-era demands. Intelligence Quotient (IQ) tests, as analyzed by Kaufman (2018), focus primarily on individual problem-solving capabilities without considering technological augmentation. These measures, developed in an era of individual cognitive work, fail to capture the collaborative and augmented nature of modern intellectual tasks.

Standard cognitive assessments typically evaluate capabilities such as pattern recognition, verbal comprehension, and mathematical reasoning in isolation from technological tools. As Dellermann et al. (2019) demonstrate, these traditional measures show limited correlation with an individual's ability to effectively leverage AI capabilities. This disconnect highlights the need for new assessment approaches that specifically address human-AI collaborative capabilities.

**2.2 Digital Literacy Measures**

Current approaches to measuring digital literacy and technological capability also prove insufficient for evaluating AI interaction skills. While frameworks exist for assessing general technological proficiency, as outlined by van Deursen and van Dijk (2016), these measures typically focus on operational skills rather than strategic collaboration capabilities. The unique cognitive demands of AI interaction require more sophisticated assessment approaches.

Digital literacy assessments often emphasize technical knowledge and tool familiarity over the meta-learning and adaptive reasoning capabilities crucial for effective AI collaboration. As Parasuraman and Wickens (2008) argue, successful human-automation interaction requires distinct cognitive capabilities that exceed traditional digital literacy measures.

**2.3 Professional Competency Frameworks**

Existing professional competency frameworks have begun to incorporate elements of technological capability but typically lack systematic approaches to evaluating AI interaction skills. Research by Autor (2015) indicates that while organizations increasingly recognize the importance of AI capabilities, they lack structured ways to assess these skills in their workforce.

Professional certification programs and skill assessments generally focus on specific technical capabilities rather than the broader cognitive skills required for effective AI collaboration. This limitation creates significant challenges for organizations attempting to evaluate workforce readiness for AI implementation.

**2.4 Educational Assessment Methods**

Current educational assessment methods show similar limitations in addressing AI-era capabilities. Traditional academic measures, as analyzed by Mislevy et al. (2017), fail to capture the complex interaction between human cognition and AI tools. This gap creates challenges for educational institutions attempting to prepare students for an AI-augmented workforce.

The rapid evolution of AI capabilities further complicates existing assessment approaches. As Hancock et al. (2020) note, successful human-AI collaboration requires continuous adaptation to evolving technological capabilities, a dynamic that traditional assessment methods struggle to capture.

These limitations of current approaches highlight the urgent need for new assessment frameworks specifically designed to evaluate human-AI collaborative capabilities. The following sections present a comprehensive framework for understanding and measuring these capabilities through the concept of Artificial Intelligence Quotient (AIQ).

## 3. Dimensions of Artificial Intelligence Quotient

The conceptualization of AIQ encompasses multiple distinct but interrelated dimensions that collectively determine an individual's capacity to effectively collaborate with AI systems. Each dimension represents a critical aspect of human-AI interaction capability, informed by current understanding of cognitive science and human-computer interaction research.

The dimensions of AIQ form an interconnected framework that captures the key aspects of human-AI collaborative capability. Figure 1 presents a comprehensive overview of these dimensions and their components.

The first fundamental dimension is Strategic AI Understanding, which encompasses an individual's ability to develop accurate mental models of AI capabilities and limitations. Drawing from Norman's (2013) work on human-computer interaction, this dimension measures how well individuals conceptualize AI systems' functionality and constraints. This understanding proves crucial for effective task delegation and problem decomposition, as highlighted by Parasuraman and Wickens (2008) in their research on human-automation interaction.

Prompt Engineering Intelligence represents the second critical dimension, focusing on the ability to effectively communicate requirements to AI systems. This capability draws from research in human communication theory (Clark, 1996) but applies specifically to human-AI interaction contexts. The dimension includes both linguistic precision and strategic thinking in query formulation, reflecting what Hernández-Orallo (2019) describes as a unique form of technological literacy.

Critical Evaluation Capability forms the third dimension, measuring an individual's ability to assess and verify AI-generated outputs. This dimension builds upon traditional critical thinking frameworks (Ennis, 2015) but incorporates specific considerations for AI-generated content. It includes the ability to detect errors, identify biases, and evaluate the reliability of AI outputs within specific contexts. As Wineburg (2018) demonstrates in his work on digital literacy, this capability becomes increasingly crucial in an AI-augmented information environment.

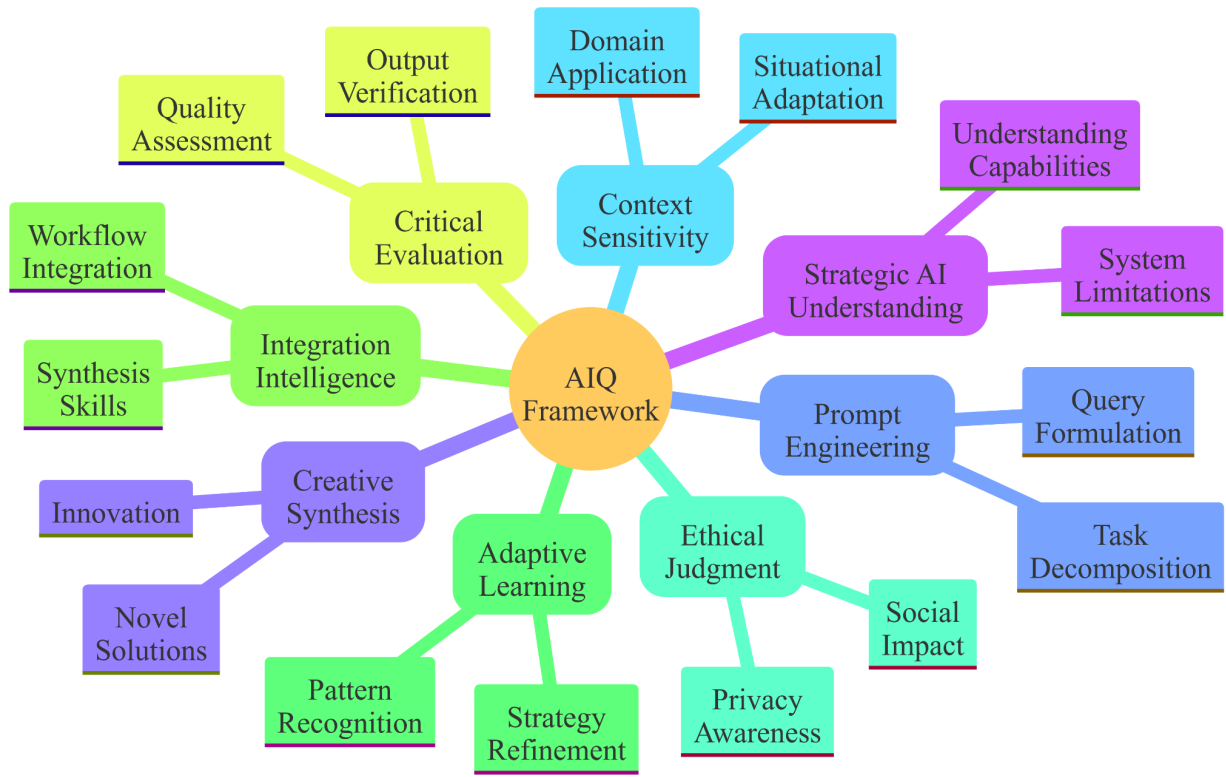

**Figure 1.** Dimensions and Components of the AIQ Framework. This mind map illustrates the eight core dimensions of Artificial Intelligence Quotient and their key subcomponents.

The fourth dimension, Integration Intelligence, measures the ability to effectively combine AI capabilities with human expertise. This dimension draws from research on distributed cognition (Hutchins, 1995) and focuses on how individuals leverage AI as a cognitive enhancement tool rather than a mere automation system. It includes the ability to identify appropriate tasks for AI delegation while maintaining human oversight of critical decisions.

Adaptive Learning Capability represents the fifth dimension, focusing on how individuals develop and refine their AI interaction strategies over time. This dimension builds upon Kolb's (1984) experiential learning theory, adapted to the specific context of human-AI interaction. It measures an individual's ability to learn from both successful and unsuccessful AI interactions, continuously improving their collaboration strategies.

The sixth dimension, Ethical Judgment in AI Utilization, addresses the crucial ability to make appropriate decisions about when and how to employ AI capabilities. This dimension incorporates elements from Vallor's (2016) work on technological virtues, measuring an individual's ability to consider ethical implications and societal impacts of AI deployment. It includes awareness of privacy concerns, bias implications, and broader societal consequences of AI utilization.

Context Sensitivity, the seventh dimension, measures an individual's ability to adapt AI utilization strategies across different situations and requirements. This dimension builds upon Sternberg's (1985) contextual intelligence framework, focusing specifically on AI application scenarios. It includes the ability to consider organizational constraints, cultural factors, and situational requirements when implementing AI solutions.

The final dimension, Creative Synthesis, evaluates an individual's ability to develop novel applications and solutions using AI capabilities. Drawing from research on creative problem-solving (Amabile, 1996), this dimension focuses on how individuals combine AI capabilities in innovative ways to address complex challenges. It measures both the originality and effectiveness of AI-enabled solution development.

These dimensions interact dynamically, creating what Anderson (2009) describes as a complex cognitive ecosystem. Success in AI collaboration typically requires competence across multiple dimensions, though individuals may show varying strength patterns across different components. Understanding these variations proves crucial for developing targeted interventions and support strategies. The implementation of AIQ assessment follows a structured process that ensures comprehensive evaluation of all dimensions while maintaining standardization and reliability. Figure 2 illustrates the systematic flow of the assessment process.

The multidimensional nature of AIQ reflects the complexity of human-AI interaction and the diverse cognitive demands it places on individuals. As AI systems continue to evolve, these dimensions may require refinement and expansion, but they provide a fundamental framework for understanding and measuring human-AI collaborative capabilities. This dimensional structure informs both assessment design and the development of interventions to enhance AI interaction capabilities.

This dimensional framework serves as the foundation for developing specific assessment methodologies and measurement tools, as discussed in subsequent sections. It provides a structured approach to understanding and evaluating the complex set of capabilities that determine success in human-AI collaboration, while maintaining flexibility to accommodate ongoing technological evolution and emerging interaction patterns.

## 4. The Case for AIQ Assessment

The rapid integration of artificial intelligence into professional and educational environments creates an urgent need for standardized assessment of human-AI collaborative capabilities. This section presents compelling arguments for the development and implementation of AIQ testing, supported by current research and emerging workplace demands.

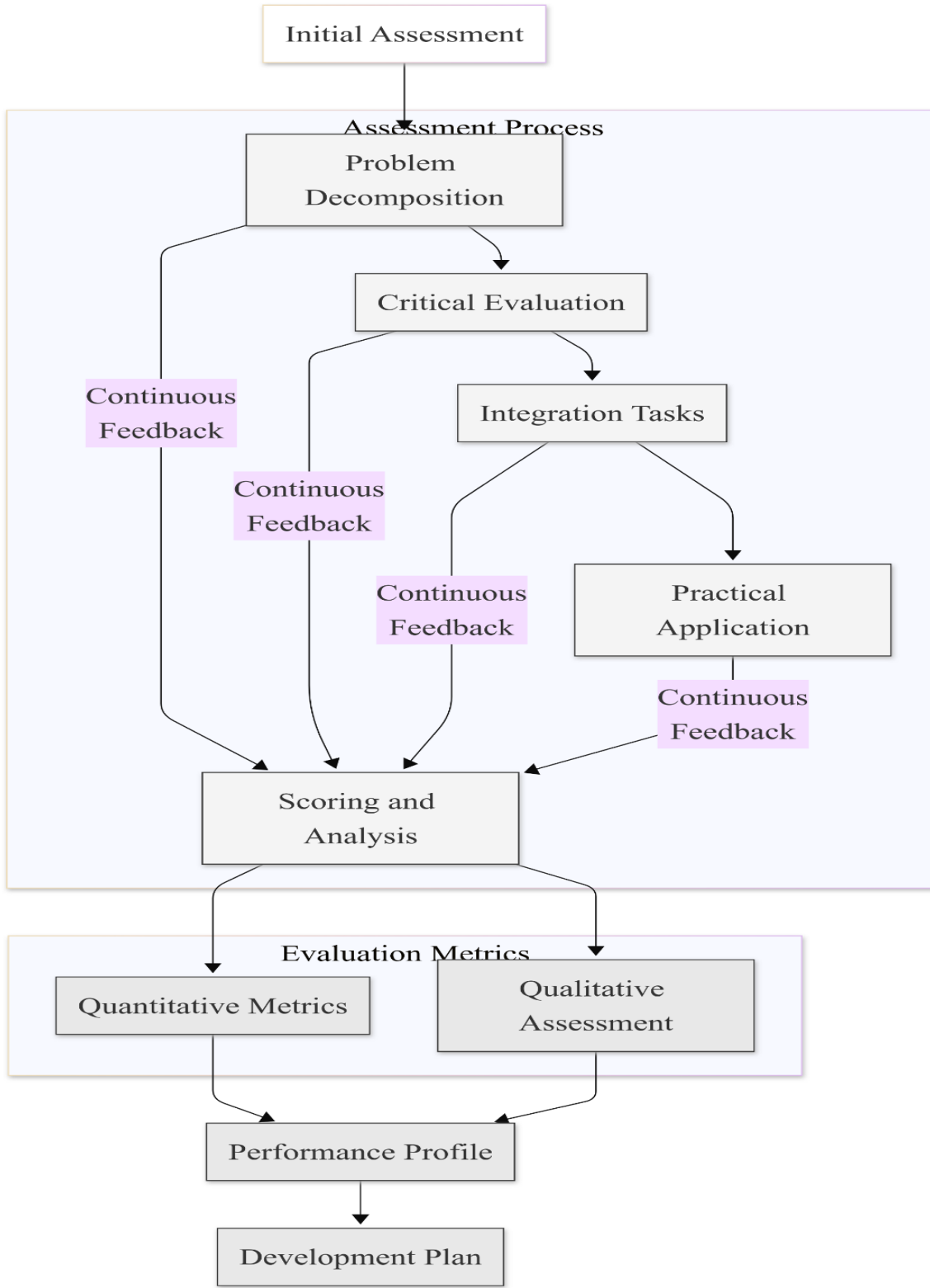

**Figure 2.** AIQ Assessment Process Flow. This diagram outlines the systematic progression of AIQ assessment, from initial evaluation through development planning, including continuous feedback and measurement components.

The traditional notion of intelligence, as measured by conventional IQ tests, fails to capture the unique cognitive demands of the AI era. As Sternberg's (2019) research on successful intelligence suggests, the ability to adapt to and leverage new technological environments represents a crucial aspect of modern cognitive capability. The emergence of AI as a fundamental tool in knowledge work necessitates new frameworks for understanding and measuring human cognitive adaptation to these technologies.

The workplace transformation driven by AI integration presents a critical argument for AIQ assessment. Studies by Brynjolfsson and McAfee (2017) demonstrate that professional success increasingly depends not just on traditional skills but on the ability to effectively collaborate with AI systems. This shift creates what Howard and Borenstein (2018) term a "cognitive augmentation imperative," where individual productivity and career advancement become increasingly tied to AI interaction capabilities.

Educational institutions face mounting pressure to prepare students for an AI-augmented workforce. Current assessment methods, however, provide limited insight into students' readiness for this new paradigm. As Hernández-Orallo (2019) argues, traditional educational metrics fail to capture the meta-learning and adaptive reasoning skills crucial for effective AI collaboration. The AIQ framework addresses this gap by providing structured evaluation of these emerging competencies.

Professional development and talent management represent another crucial domain requiring AIQ assessment. Organizations investing in AI technologies need reliable methods for evaluating employee readiness and tracking skill development. Research by Dellermann et al. (2019) indicates that successful AI implementation depends heavily on workforce capability, yet organizations lack standardized tools for measuring these capabilities.

The ethical implications of AI deployment create additional urgency for AIQ assessment. As Vallor (2016) emphasizes in her work on technological virtues, responsible AI utilization requires sophisticated judgment and ethical awareness. AIQ assessment provides a framework for evaluating these critical capabilities, helping organizations ensure responsible and effective AI implementation.

Furthermore, the democratization of AI access highlights the importance of standardized capability assessment. As AI tools become increasingly available, the digital divide may shift from access to effective utilization. Parasuraman and Wickens (2008) note that individual differences in human-automation interaction can significantly impact outcomes. AIQ assessment offers a mechanism for identifying and addressing these disparities through targeted intervention.

The economic implications of effective AI utilization provide additional motivation for AIQ assessment. Studies by Autor (2015) on technological change and labor markets suggest that the ability to complement rather than compete with AI will become increasingly valuable. AIQ assessment enables organizations to identify and develop these complementary capabilities, supporting both individual career development and organizational success.

Cognitive science research provides theoretical support for AIQ assessment. Work by Gray and Suri (2019) on human-AI collaboration demonstrates distinct patterns of cognitive adaptation that can be systematically evaluated. These findings suggest that AI interaction capability represents a measurable cognitive domain worthy of dedicated assessment tools.

The global nature of AI deployment necessitates standardized assessment approaches. As organizations increasingly operate across cultural and national boundaries, the need for consistent evaluation of AI interaction capabilities becomes more pressing. The AIQ framework provides a foundation for developing culturally adaptive assessment tools that maintain measurement validity across diverse contexts.

Finally, the rapid evolution of AI technology itself demands systematic evaluation of human adaptive capabilities. Rather than measuring static skills, AIQ assessment focuses on evaluating the meta-learning and adaptive capabilities essential for long-term success in an evolving technological landscape. This approach aligns with Hancock et al.'s (2020) research on human-automation adaptation, emphasizing the importance of measuring dynamic rather than static capabilities.

The convergence of these factors - workplace transformation, educational imperatives, professional development needs, ethical considerations, economic implications, and technological evolution - creates a compelling case for AIQ assessment. By providing a structured framework for evaluating human-AI collaborative capabilities, AIQ assessment addresses a critical gap in current measurement approaches while supporting individual and organizational adaptation to an AI-augmented future.

This framework's development represents not just an academic exercise but a practical necessity for educational institutions, organizations, and individuals navigating the transition to an AI-augmented world. The AIQ assessment framework offers a foundation for understanding and developing the cognitive capabilities that will increasingly determine success in professional and educational contexts.

## 5. Implications and Applications

The implementation of AIQ assessment carries significant implications across multiple domains, fundamentally reshaping how we approach education, professional development, and organizational strategy in an AI-augmented world.

The implementation of AIQ assessment has far-reaching implications across educational, professional, and organizational domains. Figure 3 maps the key areas of impact and application for AIQ assessment.

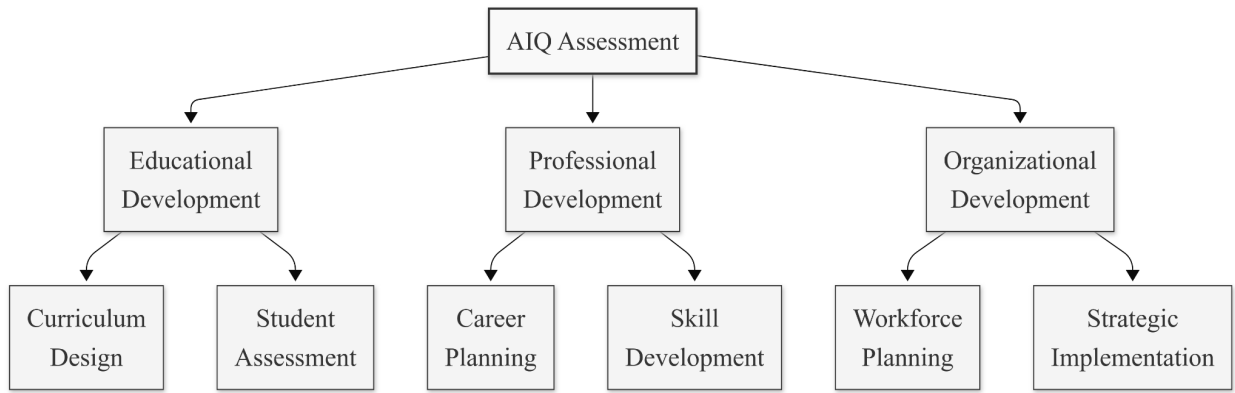

**Figure 3.** Applications and Impact Areas of AIQ Assessment. This diagram illustrates the three main domains of AIQ implementation and their specific application areas.

In educational contexts, AIQ assessment provides a framework for curriculum development and student evaluation that reflects contemporary cognitive demands. Traditional educational metrics, while valuable for measuring subject-matter knowledge, fail to capture students' readiness for AI-augmented professional environments. By incorporating AIQ assessment into educational programs, institutions can better prepare students for future workforce demands. This alignment becomes particularly crucial as Brynjolfsson and McAfee (2017) note that professional success increasingly depends on human-AI collaborative capabilities.

For organizations, AIQ assessment offers crucial insights for workforce development and talent management. As AI systems become central to professional work, organizations need reliable methods for evaluating employee capabilities and identifying development needs. AIQ assessment enables targeted training interventions and informed deployment decisions, supporting what Dellermann et al. (2019) describe as "hybrid intelligence" workflows. This capability proves particularly valuable during digital transformation initiatives, where understanding workforce readiness becomes crucial for successful AI implementation.

In professional development contexts, AIQ assessment provides individuals with structured insights into their AI collaboration capabilities. This self-awareness enables targeted skill development and career planning in an increasingly AI-augmented workplace. As Howard and Borenstein (2018) emphasize, understanding one's AI interaction capabilities becomes crucial for professional advancement and adaptation to technological change.

The implications for organizational strategy extend beyond individual assessment. AIQ frameworks enable organizations to evaluate collective capabilities, identify skills gaps, and develop targeted intervention strategies. This systematic approach to capability development supports what Parasuraman and Wickens (2008) identify as crucial factors in successful human-automation integration.

## 6. Future Directions and Challenges

The development and implementation of AIQ assessment faces several important challenges while opening new research directions and opportunities for advancement in human-AI collaboration understanding.

A primary challenge lies in maintaining assessment validity amid rapidly evolving AI capabilities. As Hernández-Orallo (2019) notes, the dynamic nature of AI technology requires assessment frameworks that can adapt to changing interaction patterns and capabilities. Future research must focus on developing flexible assessment methodologies that remain relevant despite technological advancement.

Cultural adaptability presents another significant challenge. As AI deployment crosses cultural and national boundaries, AIQ assessment must accommodate diverse perspectives and interaction patterns. Following van de Vijver's (2016) guidelines for cross-cultural assessment, future development should focus on creating culturally adaptive frameworks while maintaining measurement consistency.

Privacy and ethical considerations require careful attention in AIQ assessment development. The collection and analysis of AI interaction data must balance measurement needs with individual privacy rights. Future frameworks must incorporate what Vallor (2016) describes as "technological virtues," ensuring ethical implementation of assessment methodologies.

Research opportunities emerge in several key areas:

1. Longitudinal studies of AIQ development could provide crucial insights into how human-AI collaborative capabilities evolve over time. Understanding these developmental patterns would inform both educational interventions and professional development strategies.
2. Investigation of organizational factors influencing AIQ development could help identify environmental conditions that support or hinder effective human-AI collaboration.
3. Research into the relationship between AIQ and traditional cognitive measures could enhance our understanding of human intelligence in an AI-augmented world.
4. Studies of collective AIQ in team and organizational contexts could provide insights into effective collaborative structures for AI-augmented work environments.

The practical implementation of AIQ assessment requires ongoing refinement of measurement methodologies and scoring approaches. Future development should focus on creating efficient, scalable assessment tools that maintain measurement validity while providing actionable insights for individuals and organizations.

Additionally, the integration of AIQ assessment into existing educational and professional development frameworks presents both challenges and opportunities. Future work must address questions of standardization, certification, and integration with existing qualification systems.The evolution of AI capabilities may also require expansion of the AIQ framework to incorporate new dimensions of human-AI interaction. Continuing research should monitor emerging interaction patterns and adapt assessment methodologies accordingly.

Finally, the global implications of AIQ assessment deserve careful consideration. As AI reshapes global labor markets and educational systems, developing internationally recognized standards for AIQ assessment becomes increasingly important. This effort requires collaboration across academic, professional, and cultural boundaries to create widely applicable assessment frameworks. The future development of AIQ assessment thus requires a balanced approach that maintains scientific rigor while addressing practical implementation challenges. Success in this endeavor will significantly influence how we prepare individuals and organizations for an increasingly AI-augmented future.

## 7. Conclusion

The emergence of artificial intelligence as a fundamental tool for cognitive work creates an urgent need for new approaches to measuring and developing human capabilities. The concept of Artificial Intelligence Quotient (AIQ) addresses this need by providing a structured framework for understanding and assessing human-AI collaborative capabilities.

The compelling case for AIQ assessment rests on multiple converging factors. The transformation of professional work through AI integration, the increasing importance of human-AI collaboration in educational contexts, and the growing need for systematic workforce development all point to the necessity of standardized assessment approaches. The proposed dimensional framework for AIQ provides a foundation for developing these assessment methodologies while maintaining flexibility for future technological evolution.

As organizations and educational institutions navigate the transition to an AI-augmented future, the ability to measure and develop human-AI collaborative capabilities becomes increasingly crucial. The AIQ framework offers a structured approach to this challenge, supporting both individual development and organizational adaptation. While significant challenges remain in implementing AIQ assessment, the potential benefits for individuals, organizations, and society justify continued investment in this direction.

The future of human cognitive work will increasingly depend on effective collaboration with AI systems. By developing robust frameworks for measuring and enhancing these collaborative capabilities, we can better prepare individuals and organizations for success in an AI-augmented world. The concept of AIQ thus represents not just a measurement tool but a fundamental shift in how we understand and develop human cognitive capabilities in the age of artificial intelligence.